\documentclass[10pt,superscriptaddress,aps,pra,twocolumn,showpacs,floatfix]{revtex4-2}
\usepackage[utf8]{inputenc}
\usepackage{graphicx,amsmath,amsfonts,amssymb}
\usepackage{color}
\usepackage[colorlinks=true, allcolors={blue}]{hyperref}
\usepackage{graphicx}
\usepackage{epstopdf}
\usepackage{float}
\usepackage{placeins}
\usepackage{lipsum}
\usepackage{mathtools}
\usepackage{hyperref}
\usepackage{comment}

\usepackage{soul}

\usepackage{silence}
\newcommand{\orcid}[1]{\href{https://orcid.org/#1}{\includegraphics[width=7pt]{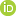}}}

\usepackage{xcolor}

\usepackage[normalem]{ulem}

\begin{document}

\preprint{APS/123-QED}

\title{Boosting State Discrimination in Quantum Brownian Motion Channel via Memory-Induced Coherence Preservation}

\author{João C. P. Porto~\orcid{0009-0006-6639-1413}}
%\email{carlosciaufpi@gmail.com}
\affiliation{Department of Physics, Federal University of Piau\'{i}, Campus Ministro Petr\^{o}nio Portela, 64049-550, Teresina, Piau\'{i}, Brazil}

\author{Pedro R. Dieguez~\orcid{0000-0002-8286-2645}}
%\email{dieguez.pr@gmail.com}
\affiliation{Department of Physics, Federal University of Paran\'{a}, P.O. Box 19044,  81531-980 Curitiba, Paran\'{a}, Brazil}

\author{Carlos H. S. Vieira~\orcid{0000-0001-7809-6215}}
%\email{vieira.carlos@ufabc.edu.br}
\affiliation{Centro de Ci\^{e}ncias Naturais e Humanas, Universidade Federal do ABC,
Avenida dos Estados 5001, 09210-580 Santo Andr\'e, S\~{a}o Paulo, Brazil.}

\affiliation{Department of Physics, State Key Laboratory of Quantum Functional Materials,
and Guangdong Basic Research Center of Excellence for Quantum Science,
Southern University of Science and Technology, Shenzhen 518055, China}

\author{Irismar G. da Paz~\orcid{0000-0002-9613-9642}}
%\email{irismarpaz@ufpi.edu.br}
\affiliation{Department of Physics, Federal University of Piau\'{i}, Campus Ministro Petr\^{o}nio Portela, 64049-550, Teresina, Piau\'{i}, Brazil}

\author{Giandomenico Palumbo~\orcid{0000-0003-1303-1247}}
%\email{giandomenico.palumbo@gmail.com}
\affiliation{CFisUC, Department of Physics, University of Coimbra, Rua Larga, 3004-516 Coimbra, Portugal}

\author{Lucas S. Marinho~\orcid{0000-0002-2923-587X}}
\email{lucas.marinho@ufpi.edu.br}
\affiliation{Department of Physics, Federal University of Piau\'{i}, Campus Ministro Petr\^{o}nio Portela, 64049-550, Teresina, Piau\'{i}, Brazil}

%date{\today}% It is always \today, today,
             %  but any date may be explicitly specified

\begin{abstract}
Preserving quantum resources in dissipative environments is a fundamental challenge in quantum information processing. While environmental interactions usually degrade quantum resources, we theoretically show that in a Quantum Brownian Motion (QBM) channel, continuous-variable state discrimination can be improved by increasing, rather than minimizing, the initial thermal noise. Specifically, without suppressing the inherent environmental dissipation, when combined with squeezing, this initial noise induces a coherence preservation mechanism driven by the transient non-thermalization of the probe with the bath. This preservation translates into a pronounced reduction in error probabilities for state discrimination between orthogonal squeezing directions. Furthermore, we also show that quadrature homodyne detection achieves near-optimal performance, approaching the Helstrom limit. These results highlight the advantage of exploiting thermal-squeezed states, offering a robust physical architecture for quantum communication in high-temperature environments.
\end{abstract}

\maketitle

%\tableofcontents

\section{Introduction}\label{sec:intro}

Originally formulated to describe the stochastic motion of particles in dissipative fluids \cite{Brown1928,Einstein1905}, the Brownian motion model has been generalized to the quantum regime, giving rise to the theory of Quantum Brownian Motion (QBM). QBM has become one of the pivotal models for the study of open quantum systems, providing fundamental insights into dissipation, decoherence, and the influence of the environment on quantum dynamics \cite{Petruccione2002book,weiss2008book,MazoBokk2008,lampo2019BrownianMotion,caldeira1983a,caldeira1983b}. 
The scope of QBM extends well beyond open-system dynamics, encompassing applications in quantum Darwinism \cite{ZurekPRL2008,ManiscalcoSciRep2016}, quantum thermometry \cite{Mirkhalaf_2024NJP,Porto2025PRA}, and quantum time crystals \cite{Tanda}. Together, these developments demonstrate the broad impact of coherence revivals on quantum information processing and complex biological media \cite{FujiiPRA2025}. Likewise, within continuous-variable systems, QBM serves as a cornerstone for the characterization of non-Markovian Gaussian dynamics~\cite{IlluminatiPRL2015}, where Gaussian interferometric power provides a quantifier of non-Markovianity \cite{LeonardoPRA2015}, and memory effects enable the operational amplification of Gaussian inputs beyond the fundamental quantum limit \cite{LeonardoPRL2017}. More recently, the QBM framework has been leveraged to engineer the degree of non-Markovianity, enabling precise control over the crossover between non-Markovian and Markovian dynamical regimes~\cite{Giovannetti2026Arxiv}.

%%%%%%%%%%%%

Among the various quantum features that can be investigated within the QBM framework, quantum coherence plays a particularly prominent role. As a fundamental resource for quantum technologies~\cite{XuPRA2016}, it has been extensively studied within the framework of quantum resource theories~\cite{Baumgratz2014PRL,PlenioReview2017,BrunnerPRL2021}. Its relevance extends to phenomena such as indefinite causal order~\cite{Dieguez2024CommPhysics}, quantum heat engines~\cite{JonasPRA2019}, and quantum parameter estimation~\cite{JonasPRA2025,Jonas2026EPJP}. In the context of QBM, quantum coherence was recently shown to exhibit a distinct form of \textit{lateral coherence} (a resurgence of coherence at intermediate times)~\cite{RoccoPRA2022}. However, existing analytical studies remain restricted to the free-motion regime, where non-Markovianity is treated perturbatively within a quasi-Markovian approximation applicable to high cutoff frequencies and elevated environmental temperatures. Consequently, the role of environmental memory beyond this regime remains largely unexplored.

In this work, we develop a theoretical framework that extends previous analytical treatments to both the low- and high-temperature regimes. This allows us to move beyond the free-motion approximation and investigate stronger non-Markovian effects, revealing a nontrivial coherence dynamics induced by high-temperature environments. Within this framework, we characterize quantum coherence from the perspective of resource theory~\cite{Baumgratz2014PRL,XuPRA2016}, establishing a direct bridge between the dynamics of open quantum systems and the operational usefulness of coherence in quantum-information protocols such as quantum state discrimination~\cite{Chefles2000,MinPRA2026}.

%%%%%%%%%%
As an application of our framework, we investigate quantum state discrimination, one of the fundamental tasks in quantum communication \cite{RevModPhysBraunstein2005,NielsenChuang}. In this protocol, information is encoded into non-orthogonal Gaussian probe states propagating through noisy continuous-variable channels, and the achievable performance is ultimately bounded by the Helstrom minimum-error criterion \cite{Helstrom1969}. We show that the short-time non-Markovian dynamics of a QBM channel~\cite{IlluminatiPRA2018} provide a remarkably favorable setting for this task. In particular, squeezing remains the more robust resource for state discrimination even as the probe's initial thermal occupation increases. More strikingly, the discrimination performance is enhanced at high probe temperatures, despite the vanishing initial purity, demonstrating that purity is not a prerequisite for operational quantum advantage in this scenario. These findings contrast sharply with previous studies, where the advantages provided by squeezing over noisy continuous-variable channels were ultimately limited by losses, allowing displacement to outperform squeezing in standard lossy channels~\cite{ParisPRA2018,Walsh2025Arxiv}. Likewise, unlike squeezed phase-shift-keyed binary discrimination under phase diffusion, whose performance strongly depends on the purity of the input state~\cite{ParisPRA2018, ParisPRL2011}, our results reveal that thermal-squeezed states become particularly effective probes when environmental memory is properly exploited.

\begin{figure}[ht]
\centering
\includegraphics[scale = 0.12]{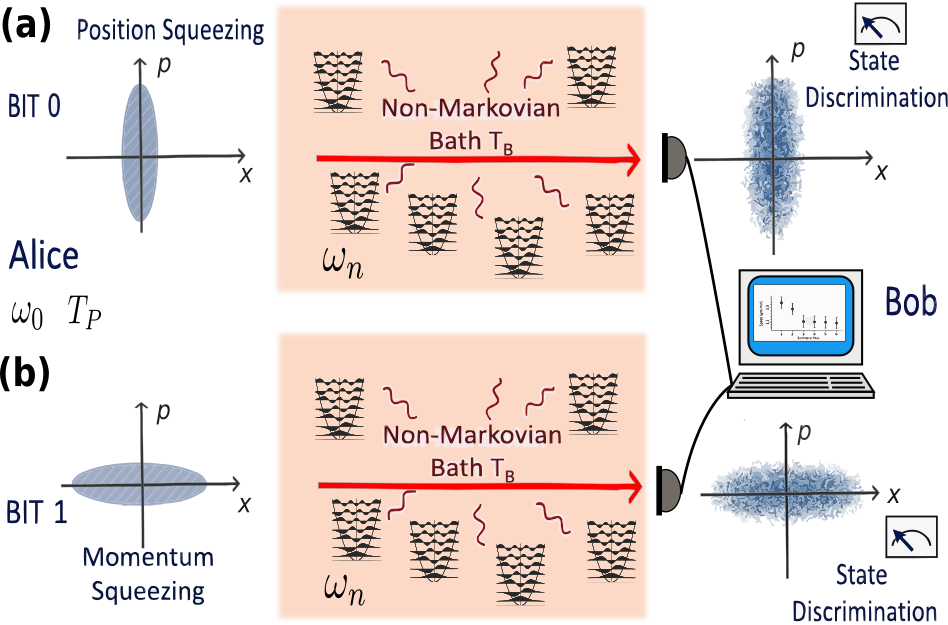}
\caption{Sketch of a quantum state discrimination protocol. First, Alice prepares a bosonic probe, $P$ (with frequency $\omega_0$ and initial temperature $T_P$), and applies squeezing along either the position quadratures (a) or momentum quadratures (b) in phase-space. For example, these distinct preparations can encode logical bits 0 and 1, respectively. She then sends these states through a non-Markovian bath at temperature $T_B$, modeled by a Quantum Brownian Motion (QBM) channel comprising an ensemble of harmonic oscillators with frequencies $\omega_n$. Finally, at the receiving end, Bob performs measurements to discriminate between these states and, by doing so, decode the logical information for communication purposes.}
\label{fig1}
\end{figure}

This work is structured as follows. In Sec.~\ref{sec: tf}, we establish the theoretical framework. This includes the non-Markovian QBM model (Sec.~\ref{sec:Gaussian_QBM}) and the formalism for quantum state discrimination (Sec.~\ref{sec:discrinimation}), where we define a Helstrom window to tightly bound the minimum error probabilities and introduce the error probability achieved from a quadrature-homodyne receiver strategy, based on variance discrimination. The quantification of quantum coherence and the entropy production rates for Gaussian states are detailed in Sec.~\ref{sec:coherence_entropy}. We discuss our results in Sec.~\ref{sec:re}, demonstrating in Sec.~\ref{sec:re_coherence} that preserving coherence can enhance state discrimination (Sec.~\ref{sec:re_discrimination}) in hot non-Markovian environments. Finally, in Sec.~\ref{sec:disc}, we summarize our findings and discuss their implications.

\section{Theoretical framework}~\label{sec: tf}

In this section, we establish the theoretical framework by first detailing the Gaussian probes and their evolution under a QBM channel, subsequently outlining the protocols for quantum state discrimination, and finally characterizing quantum coherence and entropy production.

\subsection{Gaussian States and QBM Dynamics}\label{sec:Gaussian_QBM}

Any quantum state $\hat{\rho}$ of $N$ bosonic modes is uniquely characterized by its first moment (displacement vector $\boldsymbol{\bar{r}}$) and its second moment (covariance matrix $\boldsymbol{V}$) if its Wigner representation $W(\boldsymbol{r})$ follows a Gaussian distribution~\cite{RevModPhys2012Lloyd} given by
\begin{gather}
W (\boldsymbol{r}) = \frac{\exp\left[-\frac{1}{2}(\boldsymbol{r}-\boldsymbol{\bar{r}})^{\text{T}}\boldsymbol{V}^{-1}(\boldsymbol{r}-\boldsymbol{\bar{r}})\right]}{(2\pi)^N \sqrt{\text{det}\boldsymbol{V}}}, \\ \;\;\;\;\ \boldsymbol{\bar{r}} = \langle \hat{\boldsymbol{r}} \rangle = \text{Tr}(\hat{\boldsymbol{r}}\hat{\rho}), \;\;\;\; V_{ij} = \frac{1}{2} \langle \{ \Delta\hat{r}_i, \Delta\hat{r}_j \}\rangle,
\end{gather}
where $\hat{\boldsymbol{r}} = (\hat{q}_1,\hat{p}_1,\dots,\hat{q}_N,\hat{p}_N)^{\text{T}}$ is the quadrature vector operator satisfying the canonical commutation relations $[\hat{r}_i,\hat{r}_j]=i \hbar \boldsymbol{\Omega}_{ij}$. Here, $\boldsymbol{\Omega} = \bigoplus_{k=1}^{N} \boldsymbol{\omega}$ denotes the symplectic matrix, with $\boldsymbol{\omega} = i \sigma_{y}$ being defined through the Pauli $y$-matrix~\cite{RevModPhys2012Lloyd}. Additionally, $\Delta\hat{r}_i = \hat{r}_i - \langle \hat{r}_i\rangle$, and $\{  ,  \}$ denotes the anticommutator.

 To investigate the temporal evolution of quantum coherence and its effect on the state discrimination protocol, we analyze a single-mode Gaussian probe $P$ (with frequency $\omega_0$ and initial temperature $T_P$) that interacts with a bosonic environment via a QBM channel (see Fig.~\ref{fig1}). The exact master equation, defined without relying on the Born-Markov approximation, models the interaction and dissipation processes, with coefficients determined by the bath spectral density $J(\omega)$ and bath temperature $T_B$ (see Appendix~\ref{app:QBM}). In general, the temporal evolution of the probe is fully characterized by a dynamical mapping of its moments~\cite{VasilePRA2009, IlluminatiPRA2018,Porto2025PRA}:
\begin{equation}\label{eq:gaussian_map}
\boldsymbol{\bar{r}}(t) = \boldsymbol{T}(t) \boldsymbol{\bar{r}}(0),\;\;\;\;\;\; \boldsymbol{V} (t) = \boldsymbol{T}(t) \boldsymbol{V}(0) \boldsymbol{T}^{\mathsf{T}}(t) +\boldsymbol{N}(t),
\end{equation}
where the specific dynamics are governed by the implicit time-dependence of the matrices $\boldsymbol{T}(t)$ and $\boldsymbol{N}(t)$. While this mapping is exact for arbitrary times, in our subsequent analysis, we assume weak system-reservoir coupling and specifically focus on short non-Markovian timescales. The drift matrix $\boldsymbol{T}(t)$ accounts for damping and rotation, and $\boldsymbol{N}(t)$ represents environmental noise. The explicit forms and the short-time approximation utilized for these matrices are detailed in Appendix~\ref{app:QBM}. We adopt a general single-mode Gaussian state as the initial probe configuration: $\hat{\rho}(0) = \hat{D}(\alpha) \hat{S}(\zeta) \hat{\rho}_{\mathrm{th}}(\bar{n}_{T_P}) \hat{S}^\dagger(\zeta) \hat{D}^\dagger(\alpha)$. The thermal core state $\hat{\rho}_{\mathrm{th}}(\bar{n}_{T_P})$ is characterized by the mean occupation number $\bar{n}_{T_P}= [\exp(\hbar \omega_0 / k_B T_P)-1]^{-1}$. This term describes the initial probe temperature $T_P$, which follows the Bose-Einstein statistics. The probe is also subjected to displacement ($\alpha$) and squeezing ($\zeta = r e^{i2\chi}$), where $r$ represents the amplitude and $\chi$ denotes the squeezing phase. Recently, the connection between this squeezing phase and position-momentum correlations has been explored in various applications~\cite{OzielPRA2026,PortoEPJP2026,PortoPRA2025, ThiagoNJP2025,Porto2024Scripta,MarinhoSciRep2024,MarinhoPRA2020, LustosaPRA2020,Oziel2019MPLA}. In turn, the initial moments are expressed as follows:~\cite{MarianPRA1993,Olivares2012}:
\begin{widetext}
\begin{gather}\label{eq:initial_moments}
\boldsymbol{\bar{r}}(0) = \begin{pmatrix} \bar{q}_0 \;\;\; \bar{p}_0 \end{pmatrix}, \;\;\;\;\;\;\;\;\;\;  \boldsymbol{V}(0) = (\bar{n}+1/2)\begin{pmatrix}\cosh(2r) - \sinh(2r)\cos(2\chi) & \sinh(2r)\sin(2\chi) \\ \sinh(2r)\sin(2\chi) & \cosh(2r) + \sinh(2r)\cos(2\chi)\end{pmatrix}.
\end{gather}
\end{widetext}
Here, $(\bar{q}_0=\sqrt{2}\text{Re}[\alpha], \bar{p}_0 =\sqrt{2}\text{Im}[\alpha] )$ determines the mean initial displacement vector in the phase space.

\subsection{Quantum State Discrimination}\label{sec:discrinimation}

In quantum communication, Alice encodes information using the binary logical bits 0 and 1, which are mapped to the quantum states $\hat{\rho}_1$ and $\hat{\rho}_2$, respectively. A primary task for the receiver, Bob, is to accurately distinguish between these states with minimal error as they are transmitted through a general thermal and noisy quantum channel, thus allowing him to decode the information correctly~\cite{RevModPhysBraunstein2005} [see Fig.~\ref{fig1} (right)]. Since it is generally impossible to perfectly distinguish between two non-orthogonal quantum states~\cite{NielsenChuang,Barnett2009AdvOptPhoton}, such as Gaussian states, quantum state discrimination protocols~\cite{Chefles2000,SalazarPRA2012} typically follow two well-known approaches: one focused on unambiguous state discrimination and the other on minimizing the error in state distinguishability. In the latter case, the minimum error probability is fundamentally governed by the Helstrom bound~\cite{Helstrom1969}. For states with equal a priori probabilities, this bound is given by
\begin{equation}
    p_{\text{error}}^{\text{min}} =\frac{1}{2}[1- D(\hat{\rho}_1, \hat{\rho}_2)],
\end{equation}
where $D(\hat{\rho}_1, \hat{\rho}_2)$ is the trace distance between these states. However, determining this bound analytically is often difficult, especially in continuous-variable open systems. For this reason, we address the distinguishability of quantum states via the Bures distance \cite{Dittmann, Zanardi, Palumbo}, which serves as a reliable alternative for estimating or bounding the distinguishability between states. This choice is motivated by the fact that the Bures distance satisfies inequalities involving the trace distance (the key component of the Helstrom bound) while offering a simple closed-form expression for Gaussian states. The Bures distance is defined in terms of the quantum fidelity $\mathcal{F}(\hat{\rho}_1, \hat{\rho}_2)$ as~\cite{BreuerPRA2020}: 
\begin{equation}
D_B(\hat{\rho}_1, \hat{\rho}_2) = \sqrt{2 - 2\sqrt{\mathcal{F}(\hat{\rho}_1, \hat{\rho}_2)}}.
\end{equation}  The quantum fidelity, which quantifies the overlap or similarity between two states and ranges from 0 (for orthogonal states) to 1 (for identical states), can be explicitly determined for two single-mode mixed Gaussian states, $\hat{\rho}_1$ and $\hat{\rho}_2$, using an analytical expression involving their first two moments~\cite{Jozsa1994JMO}:
\begin{gather}
\mathcal{F}(\hat{\rho}_1, \hat{\rho}_2) = \frac{2}{\sqrt{\Delta + \delta} - \sqrt{\delta}}\\
\times\exp\left[ -\frac{1}{2}(\boldsymbol{\bar{r}}_2 - \boldsymbol{\bar{r}}_1)^T (\boldsymbol{V}_1 + \boldsymbol{V}_2)^{-1} (\boldsymbol{\bar{r}}_2 - \boldsymbol{\bar{r}}_1) \right], \nonumber
\end{gather}
where $\boldsymbol{\bar{r}}_{1,2}$ represent the mean displacement vectors and $\boldsymbol{V}_{1,2}$ are the corresponding covariance matrices of the two states. The auxiliary parameters are defined as $\Delta = 4 \det(\boldsymbol{V}_1 + \boldsymbol{V}_2)$ and $\delta = 16[\det(\boldsymbol{V}_1) - 1/4][\det(\boldsymbol{V}_2) - 1/4]$. We employ this metric to quantify the state distinguishability between two Gaussian states squeezed along orthogonal directions, specifically comparing a position-squeezed state [Fig.~\ref{fig1}(a)] with a momentum-squeezed state [Fig.~\ref{fig1}(b)]. We will investigate the thermodynamic conditions under which quantum coherence can effectively boost this distinguishability, even in the presence of noisy thermal channels. Furthermore, based on the Fuchs-van de Graaf inequalities~\cite{Graaf1999}, which link trace distance to fidelity, we can derive both an upper and lower bound on the Helstrom bound ($p_{\text{error}}^{\text{min}}$) purely in terms of the Bures distance via~\cite{RevModPhys2012Lloyd}
\begin{equation}\label{eq:helstrom_bures_bounds}
p_{-} \le p_{\text{error}}^{\text{min}} \le p_{+}.
\end{equation}
where $p_{-} = \frac{1}{2} (1 - D_B\sqrt{1 - D_B^2/4})$ and $p_{+}=  \frac{1}{2} (1 - D_B^2/2)$ are the lower and upper bounds on the Helstrom limit, providing an uncertainty window to estimate it. Attaining the minimal possible error probability enables the encoded information in these states to be, for instance, used to maximize the transmission rate between a sender and a receiver in communication channels, or to enhance performance in state-distinguishability protocols~\cite{RevModPhys2012Lloyd,Walsh2025Arxiv}.

To estimate feasible measurements that can achieve the Helstrom bound, we focus on Gaussian measurements, particularly homodyne detection. This approach has demonstrated robustness to noise~\cite{ParisPRA2018} and serves as a near-optimal receiver strategy that can be readily implemented in experimental settings~\cite{ParisPRA2013}. In turn, the minimum error probability with homodyne detection and variance discrimination strategy is given by (see details in Appendix~\ref{app:homodyne}):
\begin{equation}
    p_{\text{error}}^{\text{H}} = \frac{1}{2} \left[ 1 - \text{erf}\left(\frac{q_c}{\sqrt{2}\sigma_s}\right) + \text{erf}\left(\frac{q_c}{\sqrt{2}\sigma_{as}}\right) \right],
\end{equation}
where $ q_c = \sqrt{ \frac{2 \sigma_s^2 \sigma_{as}^2 \ln(\sigma_{as}/\sigma_s)}{\sigma_{as}^2 - \sigma_s^2} }$ is a critical quadrature threshold. If the measurement quadrature result $q$ falls within the region defined by ($|q| \le q_c$), the receiver assigns the result to the position-squeezed state (with position-squeezed variance $\sigma_s$) and declares bit $0$. If the outcome falls in the regions ($|q| > q_c$), the receiver declares bit $1$, which corresponds to an anti-squeezed position variance $\sigma_{as}$.

We next briefly review the quantification of quantum coherence in Gaussian systems and discuss its role as a resource for quantum state discrimination in the present context.

\subsection{Quantum Coherence and Entropy Production}\label{sec:coherence_entropy}

Quantum coherence is a necessary condition for entanglement and other manifestations of quantum correlations~\cite{XuPRA2016}. Within the framework of resource theory, coherence is defined relative to an incoherent reference state ($\hat{\delta}$), diagonal on a chosen orthonormal basis~\cite{Baumgratz2014PRL}. For single-mode Gaussian systems, it is demonstrated that the minimal distance to the set of incoherent states $\mathcal{I}$ is achieved when a reference incoherent state $\hat{\delta}$ is a thermal state~\cite{XuPRA2016}. This results in an explicit coherence quantifier, given by:
\begin{gather}
C[\hat{\rho}(\boldsymbol{\bar{x}},\boldsymbol{V})] =  \frac{\nu-1}{2} \ln \frac{\nu-1}{2} - \frac{\nu+1}{2} \ln \frac{\nu+1}{2} \nonumber \\ + (\bar{n}+1)\ln (\bar{n}+1) - \bar{n}\ln (\bar{n}).  \label{eq:xu_coherence}
\end{gather}
Here, $\nu = 2\sqrt{\text{det}\boldsymbol{V}}$ is the symplectic eigenvalue, intrinsically linked to the purity of the Gaussian state ($\mu= \text{Tr}(\hat{\rho}^2) = 1/\nu$), where $\nu=1$ characterizes pure states. $\bar{n}= \frac{1}{2}(\text{Tr}\boldsymbol{V}+\boldsymbol{\bar{x}}^2 -1)$ is the mean excitation number, directly quantifying the mean energy of the system. Details on the formal definition of relative entropy and the foundational postulates that any valid coherence quantifier must satisfy are found in Appendix~\ref{app:Coherence}. Crucially, this coherence measure \eqref{eq:xu_coherence} is strictly monotonic: it decreases as the symplectic eigenvalue $\nu$ increases (reflecting loss of purity) and increases as the mean excitation number $\bar{n}$ increases (indicating higher energy). Recently, the role of purity in the phenomenon of coherence freezing~\cite{daSilva2024} was studied.  Furthermore, for a reference thermal state with zero displacement, the covariance matrix simplifies to $\boldsymbol{V} = (\bar{n}+1/2) \mathbb{I}$~\cite{RevModPhys2012Lloyd}, leading to the identity $\nu = 2\bar{n} + 1$. In such cases, the mean excitation number and the symplectic eigenvalue are related by $\bar{n} = (\nu - 1)/2$, resulting in a null relative entropy of coherence, $C=0$. Therefore, it is natural to define the non-thermal excitation witness as
\begin{equation}\label{eq:thermal_parameter}
\Theta = \bar{n} - \frac{\nu-1}{2},
\end{equation}
where for purely thermal states, this quantity vanishes ($\Theta=0$). Conversely, this parameter quantifies the excess excitation number beyond the thermal contribution fixed by the symplectic eigenvalue. It therefore witnesses the non-thermal, coherence-carrying component of a single-mode Gaussian state. This parameter will serve as a useful figure of merit in interpreting our results.

The von Neumann entropy quantifies the mixedness of a quantum state, capturing the loss of purity caused by its interaction with the thermal reservoir. As the entropy increases, the probe becomes progressively more mixed, reducing its ability to preserve quantum features. For a mixed Gaussian state, it is expressed as~\cite{Horhammer2008JSP,RoccoPRA2022} $
    S_{\text{vN}}(\hat{\rho})= k_B\frac{1-\mu}{2\mu}\ln \frac{1+\mu}{1-\mu} - k_B\ln \frac{2\mu}{1+\mu},$
where $\mu$ is the purity of the state. Therefore, the corresponding von Neumann entropy production rate is~\cite{RoccoPRA2022}
\begin{equation}
    \frac{dS_{\text{vN}}}{dt}= -\frac{k_B}{2\mu^2}\frac{d\mu}{dt} \; \ln \frac{1+\mu}{1-\mu}. 
\end{equation}
We investigate here the specific thermodynamic conditions under which this entropy production rate becomes negative, a signature typically associated with information backflow from the environment to the system~\cite{BreuerRevModPhysics2016}. Furthermore, we will analyze the impact of such non-Markovian dynamics on the preservation and possible revival of the probe's quantum coherence.

\section{Results}\label{sec:re}

In this section, we present our results characterizing the thermodynamic conditions under which quantum coherence enhances state discrimination.

\subsection{Quantum Coherence and Entropy Production}\label{sec:re_coherence}

\begin{figure}[htbp!]
\centering
\includegraphics[scale = 0.42]{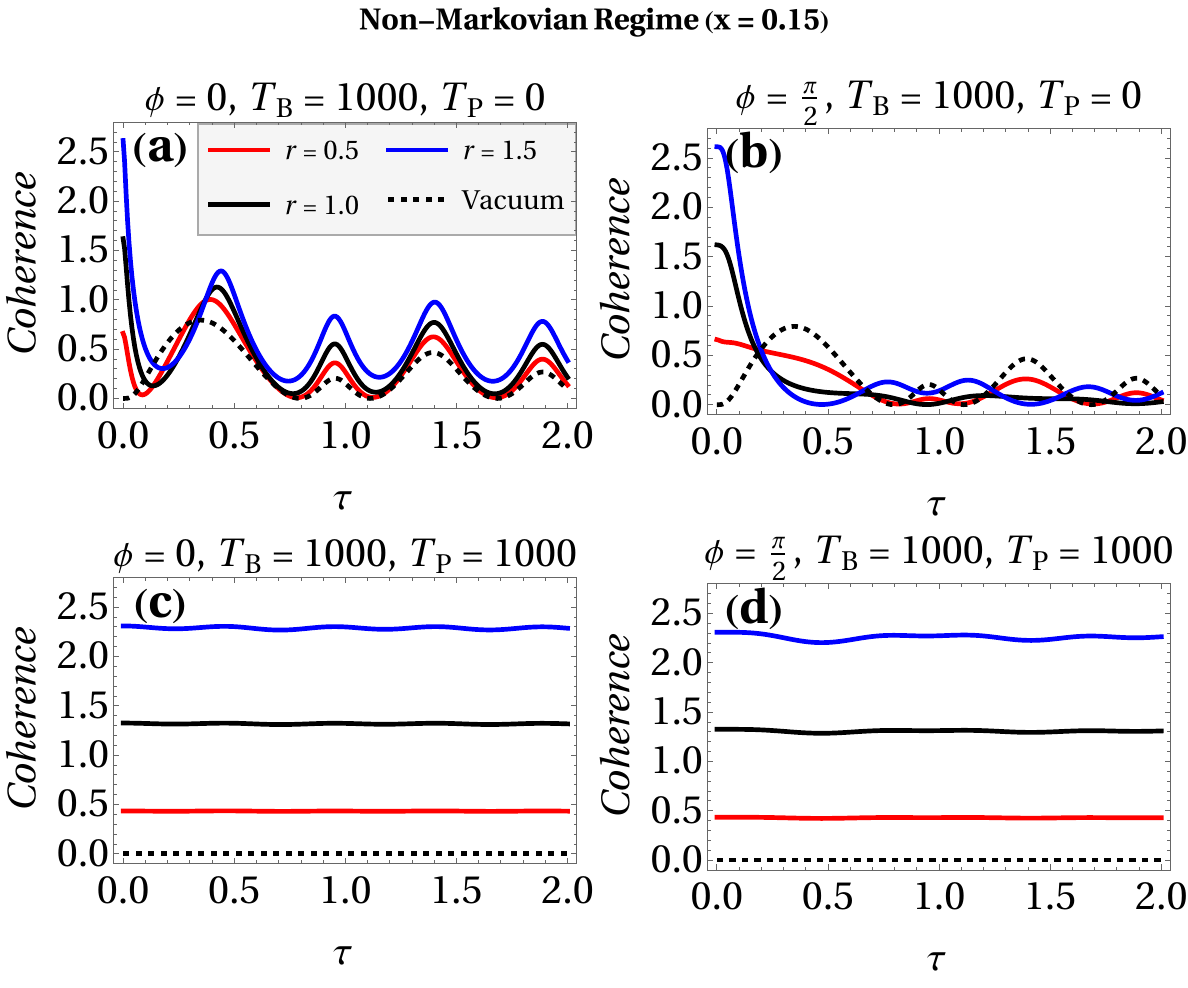}
\includegraphics[scale = 0.42]{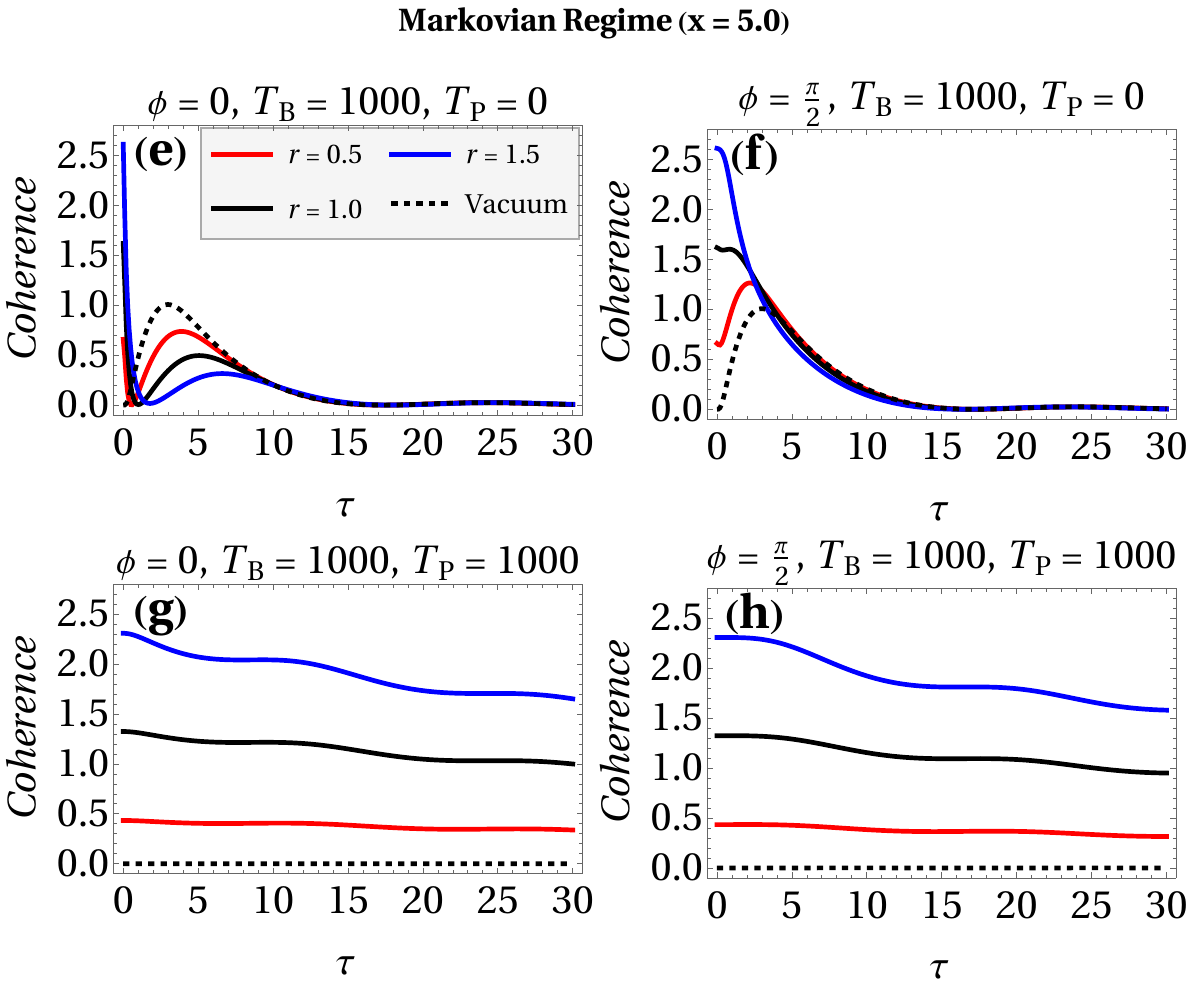}
\caption{Dynamics of the relative entropy of coherence across non-Markovian ($x = 0.15$, top block) and Markovian ($x = 5.0$, bottom block) regimes for a fixed bath temperature ($T_B = 1000$) and zero displacement ($\alpha = 0$). We contrast initial probe temperatures $T_P = 0$ [(a,b,e,f)] and $T_P = 1000$ [(c,d,g,h)], alongside position ($\phi=0$) and momentum ($\phi=\pi/2$) squeezing. Initial squeezing amplitudes are $r=0.5$ (red), $1.0$ (solid black), $1.5$ (blue), with the vacuum case ($r=0$) as a dashed black line.}
\label{fig2}
\end{figure}

Our subsequent analysis focuses specifically on the Ohmic regime (characterized by a linear system-bath interaction spectral density scaling, as detailed in Appendix~\ref{app:QBM}). Throughout this paper, we adopt natural units ($\hbar=1$, $k_B=1$) and operate in the weak coupling regime, fixing the coupling strength at $g=0.1$ and the bath temperature at $T_B=1000$. We focus on this regime to examine significant system-environment interactions and omit results for the negligible environmental effects in the zero-temperature limit. Furthermore, all physical quantities are scaled using two dimensionless parameters: time $\tau = \omega_c t$ and the non-Markovianity witness $x = \omega_c / \omega_0$~\cite{IlluminatiPRA2018}. The parameter $x$ effectively compares the correlation time scale of the environment ($\tau_E$) to the relaxation timescale ($\tau_R$), whose relation is also captured by the ratio of the natural frequency of the probe ($\omega_0$) to the cutoff frequency of the spectral density ($\omega_c$). Consequently, the limit $x \ll 1$ characterizes a regime of significant non-Markovian effects, while $x \gg 1$ denotes the approach to the Markovian limit.

\begin{figure}[htbp]
\centering
\includegraphics[scale = 0.44]{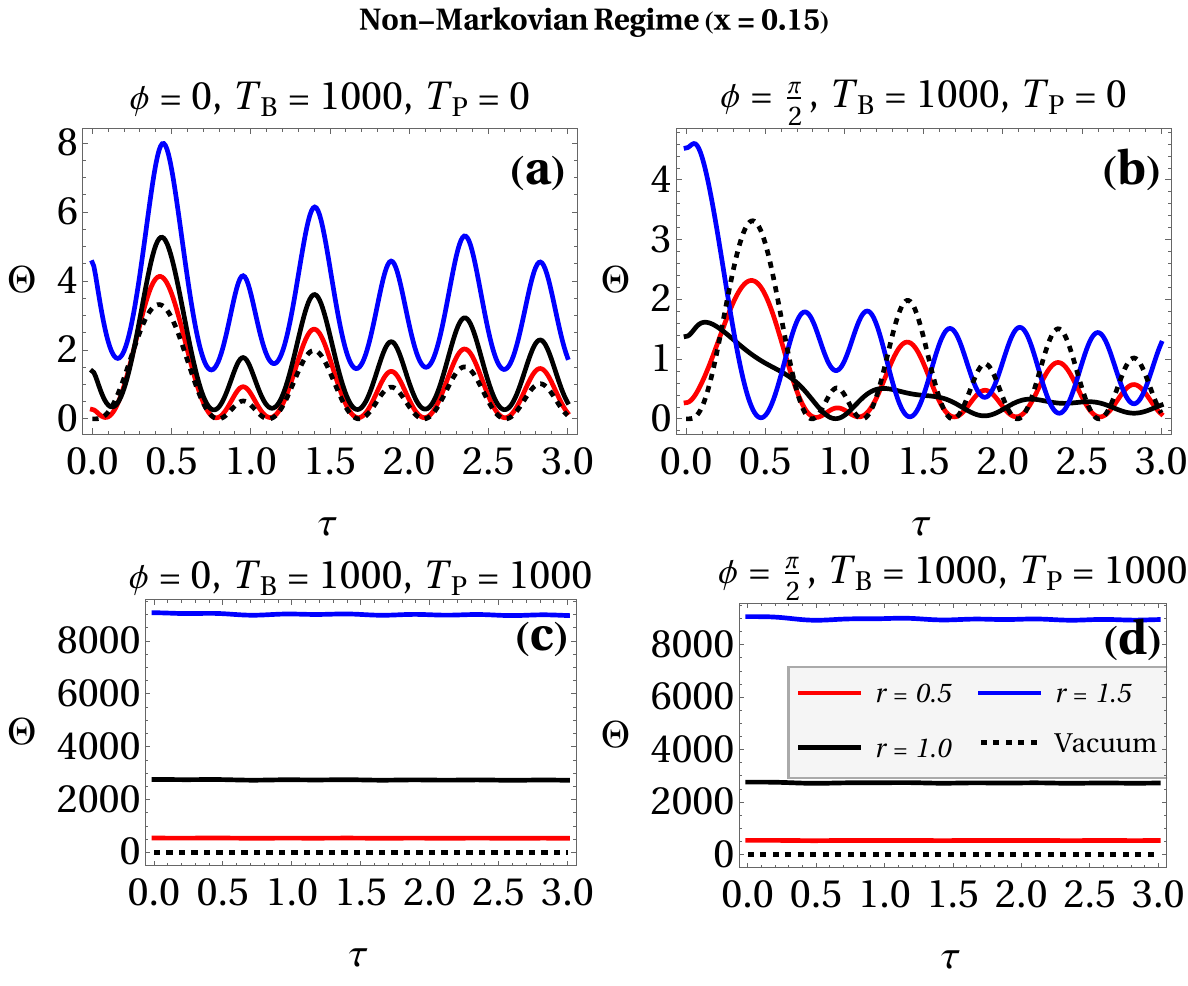}
\includegraphics[scale = 0.44]{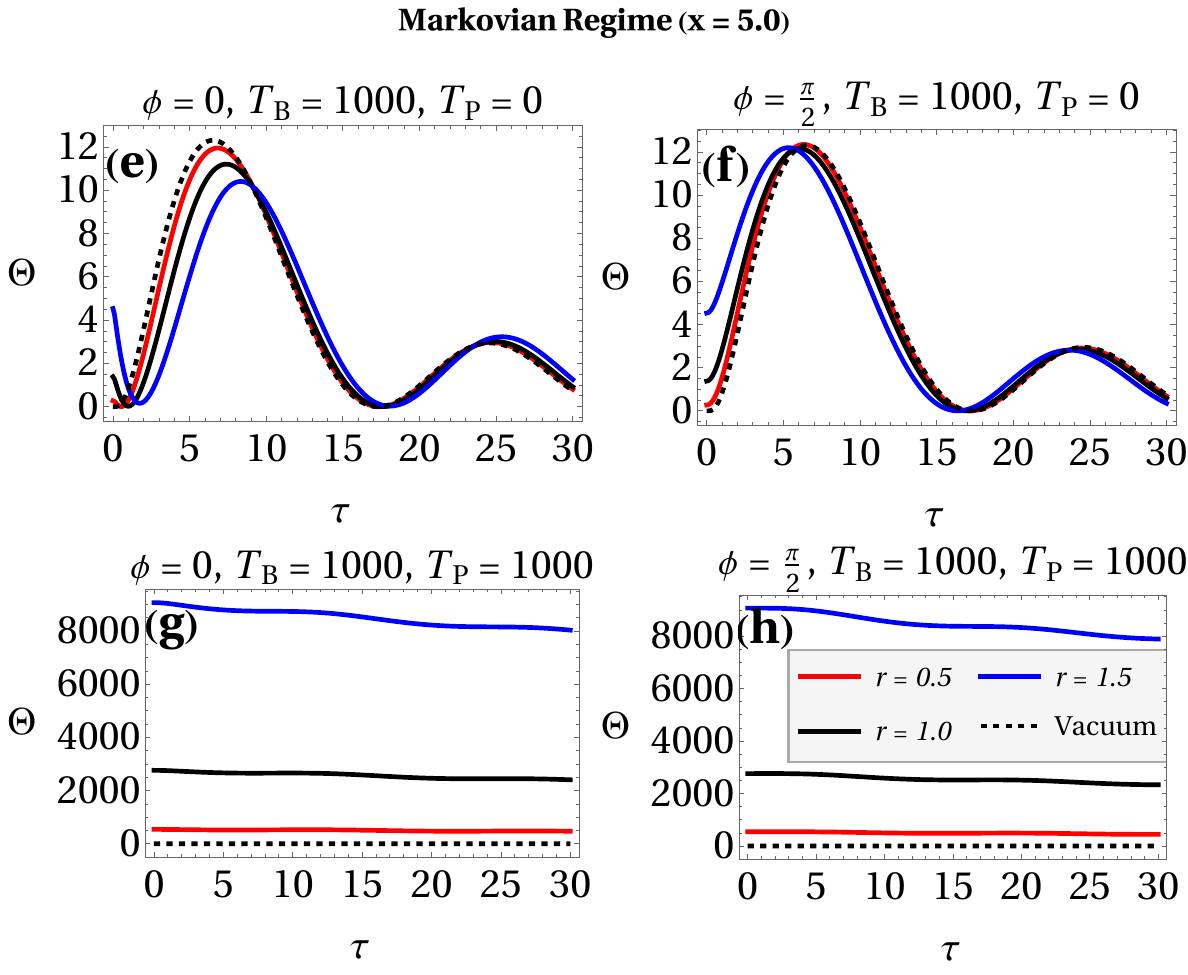}
\caption{Dynamics of the non-thermal excitation witness $\Theta$ across non-Markovian ($x = 0.15$, panels a-d) and Markovian ($x = 5.0$, panels e-h) regimes for a fixed bath temperature ($T_B = 1000$) and zero displacement ($\alpha = 0$). We contrast initial probe temperatures $T_P = 0$ [(a,b,e,f)] and $T_P = 1000$ [(c,d,g,h)], alongside position ($\phi=0$) and momentum ($\phi=\pi/2$) squeezing. Initial squeezing amplitudes are $r=0.5$ (red), $1.0$ (solid black), $1.5$ (blue), with the vacuum case ($r=0$) as a dashed black line.}
\label{fig3}
\end{figure}

\begin{figure}[htbp]
\centering
\includegraphics[scale = 0.44]{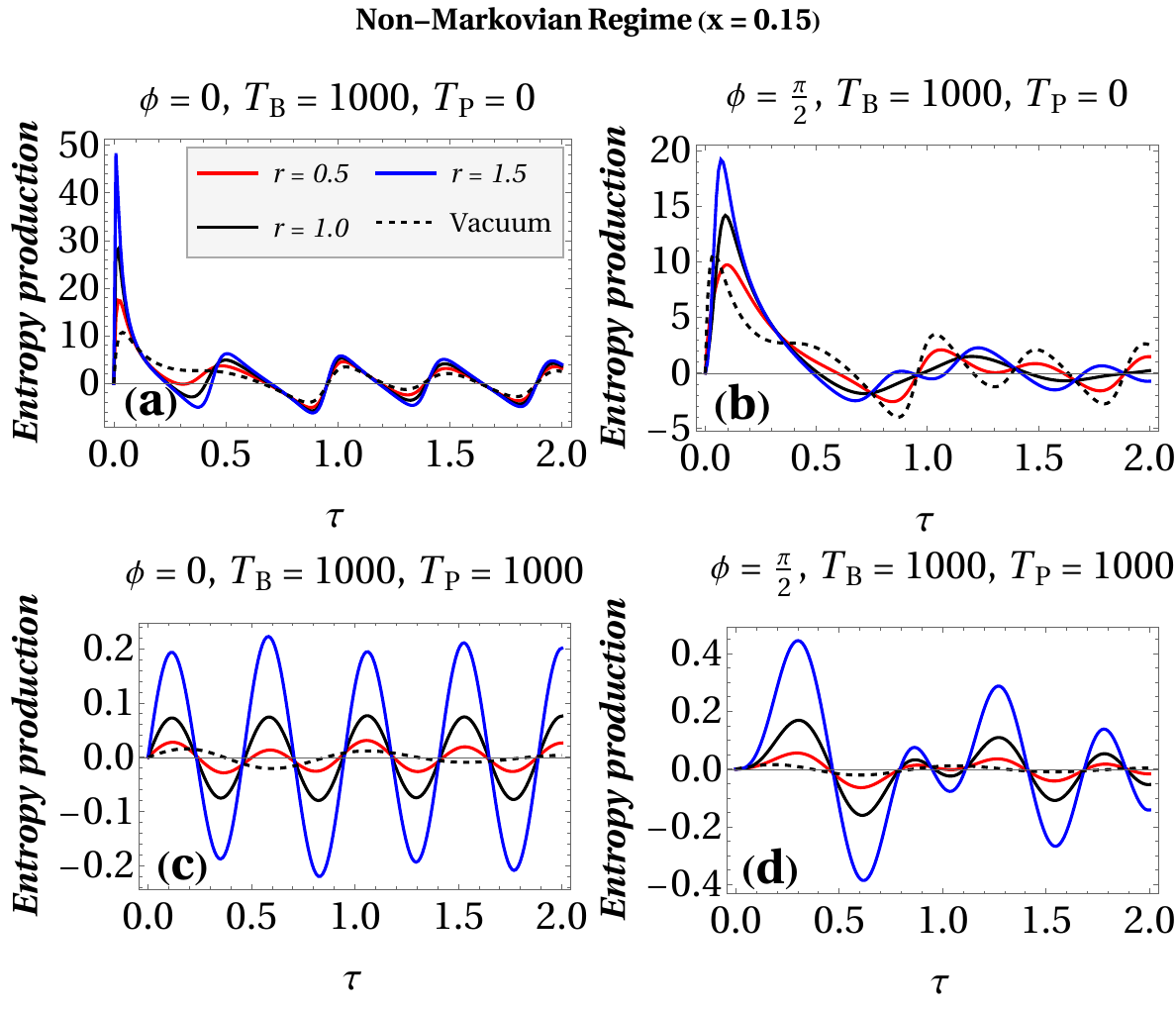}
\includegraphics[scale = 0.44]{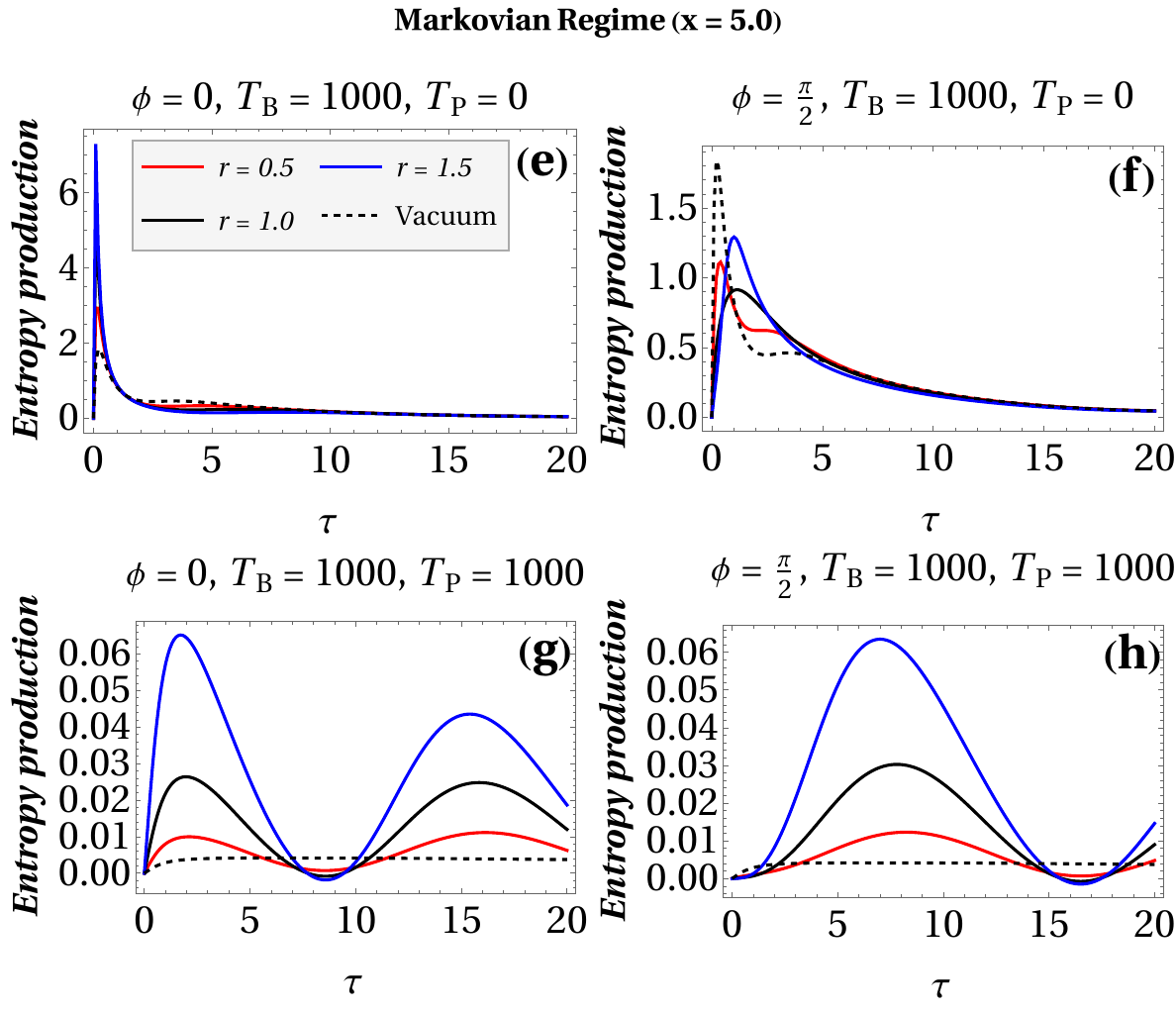}
\caption{Dynamics of the entropy production rate across non-Markovian ($x = 0.15$, panels a-d) and Markovian ($x = 5.0$, panels e-h) regimes for a fixed bath temperature ($T_B = 1000$) and zero displacement ($\alpha = 0$). We contrast initial probe temperatures $T_P = 0$ [(a,b,e,f)] and $T_P = 1000$ [(c,d,g,h)], alongside position ($\phi=0$) and momentum ($\phi=\pi/2$) squeezing. Initial squeezing amplitudes are $r=0.5$ (red), $1.0$ (solid black), $1.5$ (blue), with the vacuum case ($r=0$)  as a dashed black line.}
\label{fig4}
\end{figure}

To provide a physical insight, we first highlight the quantum coherence of our initial thermal squeezed displaced state, with first and second momenta given in~\eqref{eq:initial_moments}. The initial coherence is  determined by substituting $\nu(0) = 2\bar{n}_{T_P}+1$ and $\bar{n}(0) = (2\bar{n}_{T_P}+1)\sinh^2r + \bar{n}_{T_P}+ |\alpha|^2$ into the general coherence expression \eqref{eq:xu_coherence}. Critically, not only does the squeezing process ($r>0$) or displacement ($|\alpha|>0$) itself produce excitations, contributing to the initial mean number $\bar{n}(0)$, but the initial thermal fluctuations also contribute substantially. Interestingly, as we will demonstrate in the results, this interplay can, under certain thermodynamic conditions, preserve quantum coherence by increasing the probe thermal noise ($T_P$) when squeezing is present ($r>0$). Crucially, in the limit of vanishing initial thermal excitations ($\bar{n}_{T_P} \to 0$), we recover the well-established results of~\cite{XuPRA2016} for pure states. Notably, in this limiting pure case, the relative entropy of coherence equates to the channel capacity of an identity Gaussian channel, $\mathcal{C}= (\bar{n}+1)\ln (\bar{n}+1) - \bar{n}\ln (\bar{n})$~\cite{YuenOzawaPRL1993}. This equivalence dictates that any preservation of coherence directly manifests as an enhanced channel capacity, establishing a clear physical basis for the performance improvements in the subsequent state discrimination analysis. Furthermore, in these initial cases, the coherence measure is insensitive to both the displacement phase and the squeezing phase $\chi$. As will be demonstrated below, this characteristic changes significantly during dynamics within the QBM channel, where coherence becomes highly sensitive, in particular, to the squeezing phase.

Moreover, the initial non-thermal excitation witness \eqref{eq:thermal_parameter} (quantifying the excess non-thermal excitations prior to non-Markovian dynamics) becomes $\Theta(0)=(2\bar{n}_{T_P}+1)\sinh^2r + |\alpha|^2$. This expression reveals that the system initially possesses a non-thermal component, provided that squeezing is present ($r>0$) or displacement is applied ($|\alpha|>0$). Crucially, the probe noise ($\bar{n}_{T_P}$) not only adds to the total energy, but actually amplifies the contribution to this non-thermal component that arises from initial squeezing. Building upon this, we explore the role of these non-thermal excitations in the entropy production rate and their effects on quantum coherence across different thermodynamic and memory regimes. Our analysis will reveal a counterintuitive phenomenon: instead of merely serving as a source of degradation, increasing the initial probe thermal noise ($T_P$) can, under certain conditions, lead to a significant enhancement of quantum coherence, provided initial squeezing is present.

To systematically investigate the dynamical interplay between the probe's initial parameters (squeezing and thermal noise) and environmental memory regimes, Figures \ref{fig2}, \ref{fig3}, and \ref{fig4} collectively present the time evolution of the relative entropy of coherence, the non-thermal excitation witness, and the entropy production rate, respectively. All three figures share a unified parametric layout. We consistently contrast the strongly non-Markovian regime ($x = 0.15$) with the approaching Markovian limit ($x = 5.0$), while keeping the high-temperature bath fixed ($T_B = 1000$). Each figure isolates the effect of thermal noise from the initial probe by comparing a cold probe ($T_P = 0$) to a high-temperature probe ($T_P = 1000$). The role of initial quantum resources is captured by varying the squeezing amplitudes ($r$) and comparing their behavior for orthogonal phase orientations ($\phi = 0$ and $\phi = \pi/2$). To strictly isolate these squeezing dynamics, the initial displacement is fixed to zero ($\alpha=0$). The figures illustrate how these excess non-thermal excitations (Fig. \ref{fig3}) and the associated thermodynamic costs (Fig. \ref{fig4}) influence the generation, preservation, and dynamic evolution of quantum coherence (Fig. \ref{fig2}) in open dynamics. Notably, when combined with initial squeezing, thermal noise preserves the relative entropy of coherence across both Markovian and non-Markovian regimes [see Figs.~\ref{fig2}(c), \ref{fig2}(d), \ref{fig2}(g), and \ref{fig2}(h)]. As illustrated in Figure~\ref{fig3}, this enhancement is driven by a pronounced non-thermal component in the transient dynamics. Furthermore, in the high-probe-temperature regime, the entropy production rate is effectively suppressed to a time-averaged near-zero value. In the next section, we exploit the robustness of these thermal-squeezed states against thermal degradation to perform state discrimination. A comparative analysis of purely displaced states ($\alpha \neq 0$), with null squeezing ($r=0$), is provided in Appendix~\ref{app:Coherence}, which demonstrates that equivalent robustness is not achieved under high initial probe thermal noise.

\subsection{Quantum State Discrimination Application}\label{sec:re_discrimination}

\begin{figure}[htpb!]
\centering
\includegraphics[scale = 0.44]{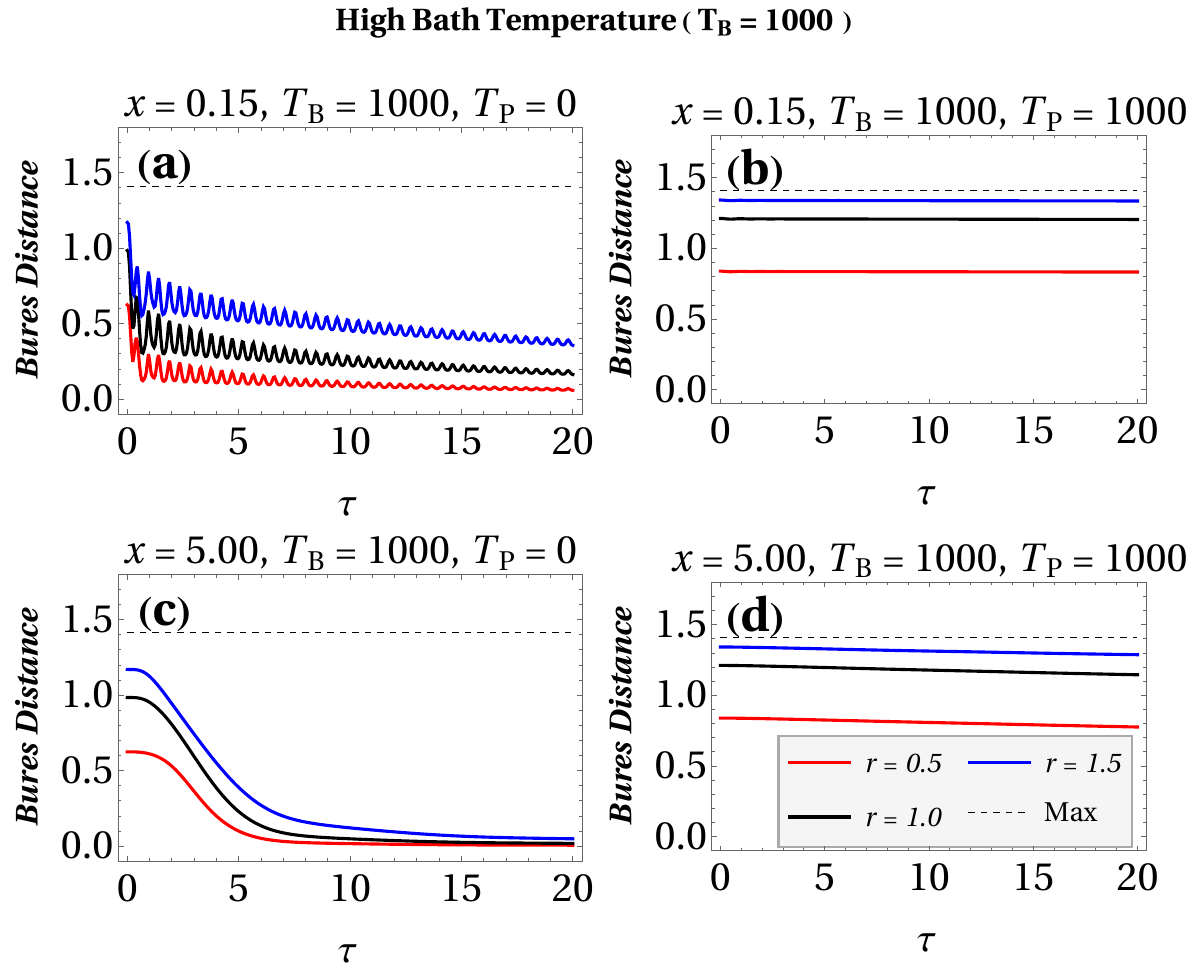}
\caption{Dynamics of the Bures distance for variance-encoded Gaussian states ($\phi=0$ versus $\phi=\pi/2$) across non-Markovian ($x = 0.15$, panels a,b) and Markovian ($x = 5.0$, panels c,d) regimes for a fixed bath temperature ($T_B = 1000$) and zero displacement ($\alpha = 0$). We contrast initial probe temperatures $T_P = 0$ (left panels) and $T_P = 1000$ (right panels). Initial squeezing amplitudes are $r=0.5$ (red), $1.0$ (solid black), and $1.5$ (blue), with the maximum possible Bures distance ($\sqrt{2}$) marked by a dashed black line.}
\label{fig5}
\end{figure}

\begin{figure}[htpb!]
\centering
\includegraphics[scale = 0.44]{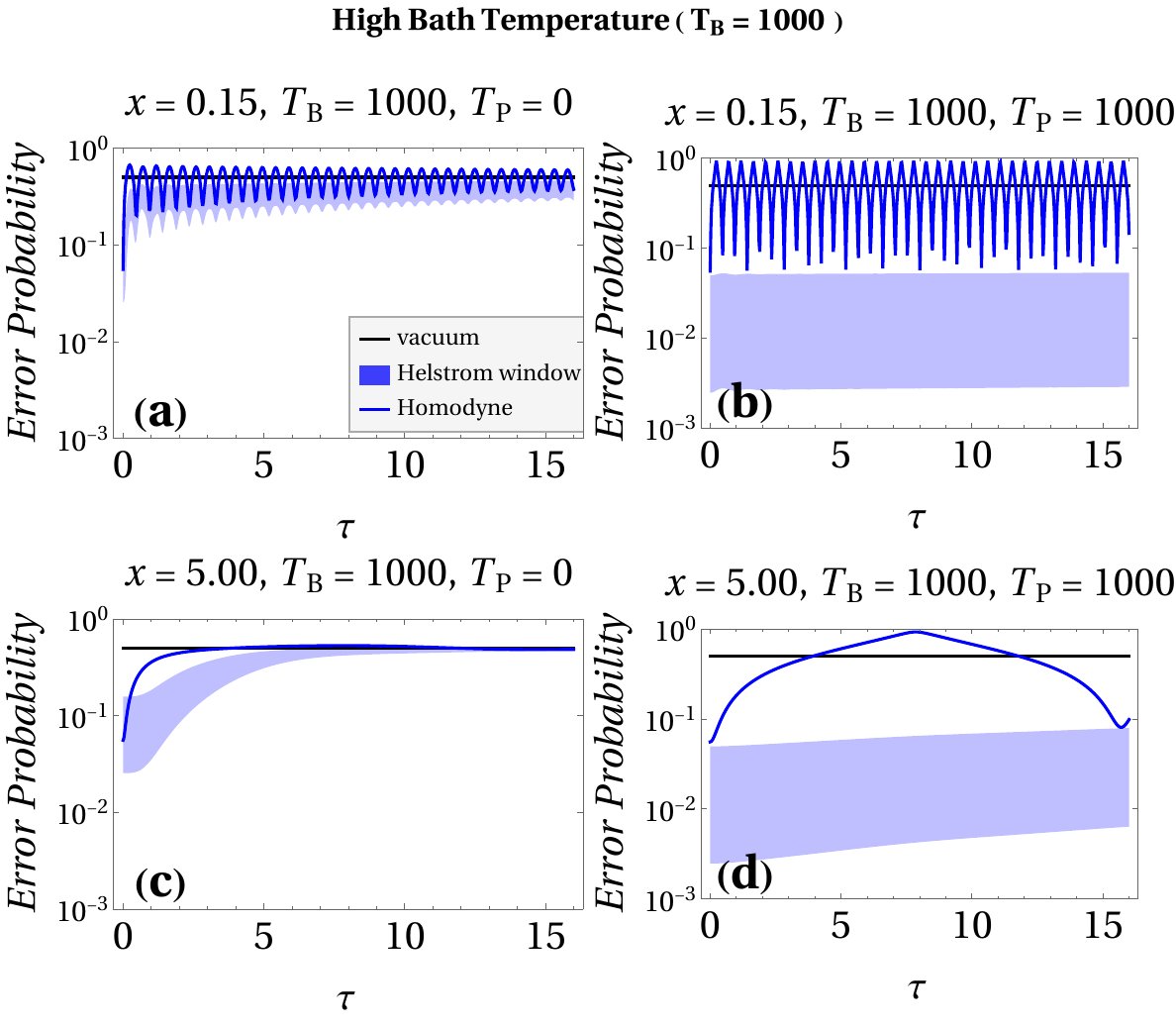}
\caption{Error probabilities for discriminating between orthogonal squeezing directions ($\phi=0$ versus $\phi=\pi/2$) in variance-encoded Gaussian states across non-Markovian ($x = 0.15$, panels a,b) and Markovian ($x = 5.0$, panels c,d) regimes for a fixed bath temperature ($T_B = 1000$) and zero displacement ($\alpha = 0$). We contrast initial probe temperatures $T_P = 0$ (left panels) and $T_P = 1000$ (right panels). The standard classical limit ($r=0$) is marked by a solid black line at $0.5$. For high squeezing ($r=1.5$), the fundamental uncertainty window bounded by Helstrom limits ($p_\pm$) is shaded blue, with the solid blue line denoting the exact error probability via optimal position-quadrature homodyne detection.}
\label{fig6}
\end{figure}

\begin{figure}[htpb!]
\centering
\includegraphics[scale = 0.44]{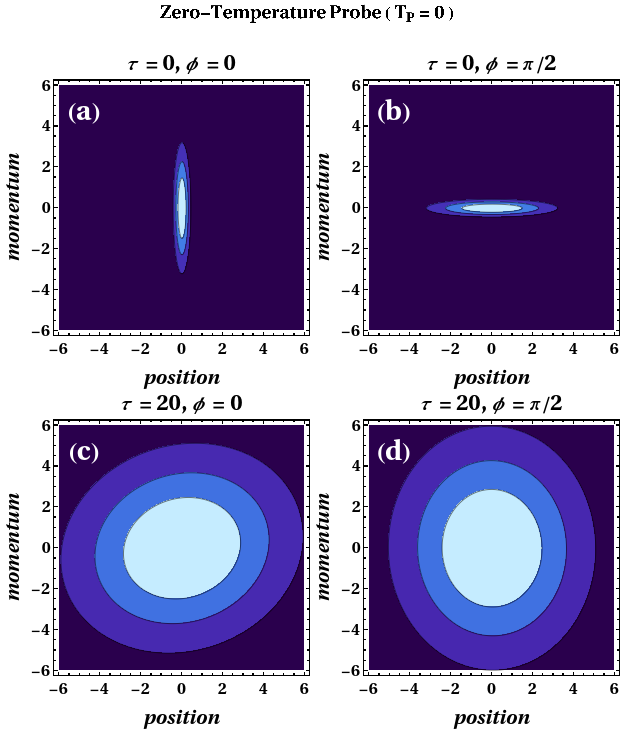}
\includegraphics[scale = 0.46]{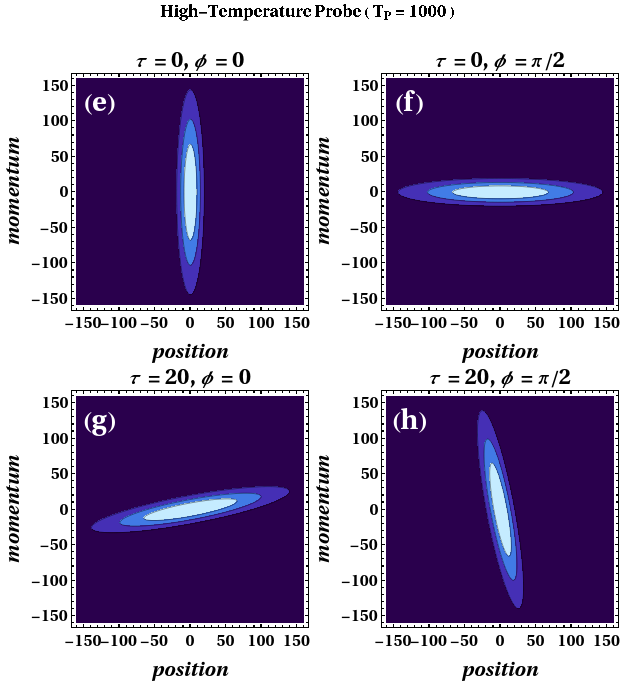}
\caption{Phase-space evolution of the Wigner function for a squeezed probe ($r=1.0$) in the non-Markovian regime ($x=0.15$) with a high-temperature bath ($T_B=1000$). We contrast initial probe temperatures $T_P=0$ (panels a-d) and $T_P=1000$ (panels e-h), comparing the initial state ($\tau=0$, top panels) to the evolved state ($\tau=20$, bottom panels) alongside position ($\phi=0$, left columns) and momentum ($\phi=\pi/2$, right columns) initial squeezing. (Note: To account for thermal expansion, the axes in panels e-h span an order of magnitude larger than those in a-d). This illustrates how increasing initial thermal noise can preserve the distinguishability between orthogonal squeezing directions.}
\label{fig7}
\end{figure}

Building upon our previous analysis, Figures~\ref{fig5} and \ref{fig6} evaluate the operational success of the phase-squeezing-encoded protocol by analyzing state distinguishability (Bures distance) and the corresponding state discrimination error probabilities. When comparing the cold and high-temperature probe regimes, the combination of initial squeezing ($r > 0$) and probe thermal noise ($T_P = 1000$) substantially increases the state distance [Fig.\ref{fig5}(b) and (d)]. Consequently, this amplified distinguishability translates into a pronounced reduction in the error probability [Fig.~\ref{fig6}(b) and (d)], enabling quadrature homodyne detection to achieve near-optimal performance approaching the Helstrom uncertainty window, attaining error probabilities on the order of 4\% at specific optimal measurement times. We note that the homodyne error exhibits strong oscillatory behavior, intermittently approaching the classical limit before returning to the 4\% minimum. This dynamic necessitates precise measurement timing to exploit the near-optimal windows. This demonstrates that initial thermal noise and squeezing serve as a powerful resource for protecting state discrimination in hot environments.

To visualize the mechanisms underlying this enhanced distinguishability, Figure \ref{fig7} presents the dynamics of the phase-space representation (Wigner function) for a squeezed probe ($r = 1.0$) interacting with a high-temperature bath ($T_B = 1000$) in the non-Markovian regime ($x = 0.15$). Mapping the states encoded with orthogonal squeezing directions ($\phi = 0$ and $\phi = \pi/2$) allows for a geometric evaluation of their distinguishability. For an initially cold probe ($T_P = 0$), the environmental interaction rapidly diffuses the Wigner distributions, suppressing the initial squeezing and causing the states with orthogonal squeezing phases to significantly overlap by $\tau = 20$. Conversely, when the probe is initialized with high thermal noise ($T_P = 1000$), the initial Wigner functions occupy a substantially larger phase-space volume. During evolution, the interplay between initial thermal expansion and squeezing dynamics preserves the distributions' orthogonal orientation, thereby maintaining state distinguishability and shielding the system from environmental degradation.

\section{Discussion}\label{sec:disc}

In this work, we analyzed the behavior of the relative entropy of coherence under different thermodynamic conditions and environmental memory regimes within a QBM framework. We found that the combined effects of initial thermal noise and squeezing support the preservation of coherence despite the noisy channel. Furthermore, although the initial coherence is independent of both the squeezing and displacement phases, as emphasized in~\cite{XuPRA2016}, the subsequent open-system dynamics of quantum coherence prove in particular to be highly sensitive to the squeezing phase. We also investigated the physical origin of this coherence preservation, linking it to the sustained presence of a non-thermal component and to the mitigated entropy production rate of the thermal bath during the transient dynamics.

Motivated by these findings, we evaluated the application of this resource (coherence) within state discrimination protocols. We demonstrated that the distance between states squeezed along mutually orthogonal axes is remarkably preserved, even under severe dissipation and excess thermal noise. This geometric resilience highlights the practical viability of utilizing not only the squeezing amplitude but also its associated phase to encode quantum information, for instance, by assigning orthogonal squeezing phases to represent the logical bits $0$ and $1$. Consequently, these phase-encoded thermal-squeezed states prove to be highly robust carriers of information across inherently noisy environments.

Although recent work has shown that displacement can outperform squeezing as a resource in the presence of loss \cite{Walsh2025Arxiv}, our system exhibits a distinct behavior during short-time non-Markovian dynamics. Specifically, we demonstrate that squeezing remains more robust for state discrimination, particularly as the initial thermal noise in the probe increases (see Appendix~\ref{app:Coherence}). Furthermore, our results show that initial-state purity is not a prerequisite for operational success in this framework. In contrast, discrimination performance is actually enhanced at high probe temperatures, where the initial purity is vanishingly small. This strict independence from initial purity contrasts sharply with the findings reported in Ref.~\cite{ParisPRA2018} for squeezing phase-shift-keyed binary discrimination in noisy channels subject to phase diffusion \cite{ParisPRL2011}, highlighting the unique advantages of exploiting thermal-squeezed states in such dissipative environments.

From an experimental perspective, optical parametric oscillators (OPOs) \cite{MillerPRL1965,Walsh2025Arxiv,BrenoPRA2024} provide an ideal platform for generating squeezing resources by exploiting nonlinear processes—such as four-wave mixing in hot alkali vapors \cite{MartinelliPRL2022} and cold-atom mirrorless OPOs \cite{FelintoPRR2026,TabosaPRA2026}. Furthermore, on the detection side, homodyne detection is a widely used technique to measure the quantum properties of light beams~\cite{MartinelliPRA2013}.

\begin{acknowledgments}
J.C.P.P. acknowledges Fundação de Amparo à Pesquisa do Estado do Piauí (FAPEPI) for financial support. P.R.D. acknowledges support from CNPq/MCTI/FNDTC 22/2024 (No. 446775/2024-0). C.H.S.V. acknowledges support from the São Paulo Research Foundation (FAPESP, Grants No. 2023/13362-0 and 2025/14546-2) and the Southern University of Science and Technology (SUSTech) for providing workspace during the research visit. I.G.P. acknowledges Grant No. 306528/2023-1 from CNPq. L.S.M. acknowledges support from the National Institute of Science and Technology on National Institute of Photonics (INFO) CNPq - INCT grant 409174/2024-6. G. P. acknowledges support from national funds by FCT - Fundação para a Ciência e Tecnologia, I.P. in the framework of the project UID/04564/2025, with DOI identifier 10.54499/UID/04564/2025.
\end{acknowledgments}

\appendix

\section{QBM Model}\label{app:QBM}

The temporal evolution of the reduced density matrix $\hat{\rho}(t)$ for a single harmonic oscillator of frequency $\omega_0$ interacting with $N$ bosonic oscillators is governed by the exact master equation~\cite{JPPaz1992PRD,IlluminatiPRA2018}:
\begin{gather}
\frac{d \rho (t)}{dt}  = - \frac{i}{\hbar} [H_S,\rho (t)] - \Delta(t) [\hat{q}, [\hat{q},\rho (t)]] \nonumber \\
+ \Pi(t) [\hat{q}, [\hat{p},\rho (t)]] - i\Gamma(t) [\hat{q}, [\hat{p},\rho (t)]], \label{eq:master_QBM}
\end{gather}
where $\Gamma(t)$ is the dissipation coefficient, while $\Delta(t)$ and $\Pi(t)$ are the normal and anomalous diffusion coefficients. In the weak-coupling regime ($g \ll 1$), these coefficients are determined by the spectral density $J(\omega)$ and the thermal occupation $N(\omega)=[\exp(\hbar\omega/k_BT_B)-1]^{-1}$ of the bath at temperature $T_B$~\cite{intravaia2003PRA}:
\begin{equation}
\Gamma (t) = g^2 \int_{0}^{t} \int_{0}^{+ \infty} d t' d\omega J(\omega) \sin\omega t' \sin(\omega_0 t'),
\end{equation}

\begin{equation}
\Delta (t)= g^2  \int_{0}^{t} \int_{0}^{+ \infty} d t' d\omega J(\omega) [2N(\omega)+1] \cos\omega t' \cos\omega_0 t',
\end{equation}
\begin{equation}
\Pi (t) = g^2  \int_{0}^{t} \int_{0}^{+ \infty} d t' d\omega J(\omega) [2N(\omega)+1] \cos\omega t' \sin\omega_0 t',
\end{equation}
Here, we utilize an Ohmic-like class of spectral densities to phenomenologically model the bath interaction in the continuum limit~\cite{Petruccione2002book,BreuerPRA2020}: $J_s(\omega) = ( \frac{\omega}{\omega_c} )^s e^{-\omega/\omega_c }$. The cases $s < 1$, $s=1$, and $s > 1$ correspond to sub-Ohmic, Ohmic, and super-Ohmic regimes, respectively, regularized by the cutoff frequency $\omega_c$. Closed expressions of the time-dependent coefficients of the master equation can be obtained in the high and low-temperature regime, i.e., for $[2N(\omega)+1] =\coth \Big( \frac{\hbar \omega}{2 k_B T} \Big) \approx \frac{2 k_B T }{\hbar \omega} $ and $[2N(\omega)+1] = \coth \Big( \frac{\hbar \omega}{2 k_B T} \Big) \approx 1 +2\exp{\Big( -\frac{\hbar \omega}{k_B T} \Big)}$, respectively (for more details, see the Appendix of Ref. \cite{VasilePRA2009}).

The transformation matrices $\boldsymbol{T}(t)$ and $\boldsymbol{N}(t)$ governing the dynamic map~\eqref{eq:gaussian_map} are defined as~\cite{IlluminatiPRA2018}:
\begin{gather}
 \boldsymbol{T}(t) = e^{-[\widetilde{\Gamma} (t)/2]} \boldsymbol{R}(t),\\
 \boldsymbol{N}(t) = [\boldsymbol{R}^{-1}(t)]^{\mathsf{T}}\Big[e^{-\widetilde{\Gamma} (t)} \int_{0}^{t} d t' e^{\widetilde{\Gamma} (t')} \boldsymbol{R}^{\mathsf{T}}(t') \boldsymbol{M}(t') \boldsymbol{R}(t')\Big] \nonumber \\ \times \boldsymbol{R}^{-1}(t), \label{eq:X_Y_app}
\end{gather}
with the rotation matrix $\boldsymbol{R}(t)$ and memory matrix $\boldsymbol{M}(t)$ defined as:
\begin{gather}\label{eq:M_R_matrices}
 \boldsymbol{R}(t)=\left(\begin{array}{cc}
\cos \omega_0 t & \sin \omega_0 t  \\
-\sin \omega_0 t   & \cos \omega_0 t
\end{array}\right), \nonumber \\ \boldsymbol{M}(t)=\left(\begin{array}{cc}
\Delta(t) & -\Pi(t)/2  \\
-\Pi(t)/2 & 0
\end{array}\right),   \; \;\widetilde{\Gamma} (t) = 2 \int_{0}^{t} \Gamma (t') dt'.
\end{gather}
In our weak-coupling, non-Markovian short-time analysis, we retain only the first term of the Taylor expansion of the integral in Eq.~\eqref{eq:X_Y_app}, neglecting higher-order terms $\mathcal{O}(g^4)$~\cite{VasilePRA2009}:
\begin{gather}
 \Big[e^{-\widetilde{\Gamma}(t)} \int_{0}^{t} d t' e^{\widetilde{\Gamma} (t')} \boldsymbol{R}^{\mathsf{T}}(t') \boldsymbol{M}(t') \boldsymbol{R}(t') \Big] \approx \nonumber \\ \int_{0}^{t} d t' \boldsymbol{R}^{\mathsf{T}}(t') \boldsymbol{M}(t')  \boldsymbol{R}(t'). \label{eq:approx_N_app}
\end{gather}
This approximation relies on the exponential factors remaining near unity, requiring $\widetilde{\Gamma}(t) \ll 1$, a condition naturally satisfied for timescales $t \ll \Gamma^{-1}$. In our chosen parameter regimes, the typical dissipation rate is on the order of $\Gamma \sim 10^{-4}$. Consequently, for our maximum evaluated time of $\tau = 30$, we safely remain in the regime where $\widetilde{\Gamma}(t) \ll 1$, thereby fully justifying its application to the transient dynamics explored here. Furthermore, to strictly guarantee the physical validity of the state under this truncation, we independently verified that the symplectic eigenvalue $\nu(\tau) = 2\sqrt{\det \boldsymbol{V}}$ never violates the fundamental physicality bound $\nu \geq 1$ at any point during the evaluated time window.

\begin{widetext}

\section{TIME-DEPENDENT COEFFICIENTS}\label{Ap:Coeff}

Here, we provide analytic expressions of the time-dependent coefficients of the master equation~\eqref{eq:master_QBM}. For simplicity, we examine only the Ohmic reservoir spectral density ($s=1$) and calculate the temperature-independent damping coefficient $\Gamma(t)$, and the diffusion coefficients in the high-temperature regime $\Delta_{T_B} (t)$ and $\Pi_{T_B} (t)$.  In the following, we write these results as follows
\begin{equation}
    \Gamma(\tau,x) = \frac{g^2}{4 x} \Bigg\{ i e^{-1/x} \Big[ \text{Ei} \Big ( \frac{1-i\tau}{x} \Big) - \text{Ei} \Big ( \frac{1+i\tau}{x} \Big) \Big] +   e^{1/x} \Big[ 2\pi + i \text{Ei} \Big ( \frac{i\tau-1}{x} \Big) - i \text{Ei} \Big (- \frac{1+i\tau}{x} \Big) \Big] - \frac{4 x\sin (\tau/x)}{1+\tau^2} \Bigg\}, 
\end{equation}
\begin{equation}
    \Delta_{T_{B}} (\tau,x,T_B) = \frac{g^2 k_B T_B e^{-1/x}}{2\hbar \omega_c} \Bigg\{  i \Big [ \text{Ei} \Big ( \frac{1- i\tau}{x} \Big) - \text{Ei} \Big ( \frac{1+ i\tau}{x}\Big)   \Big]  + e^{2/x}\Big[  2\pi +i \text{Ei} \Big ( \frac{i\tau-1}{x}\Big) - i \text{Ei} \Big ( - \frac{i\tau+1}{x}\Big)  \Big]   \Bigg\},
\end{equation}
\begin{equation}
    \Pi_{T_{B}}(\tau,x,T_B) = \frac{g^2 k_B T_B e^{-1/x}}{2\hbar \omega_c} \Bigg\{   -\text{Ei} \Big ( \frac{1- i\tau}{x} \Big) - \text{Ei} \Big ( \frac{1+ i\tau}{x}\Big) + 2\text{Ei} \Big (\frac{1}{x}\Big)     + e^{2/x}\Big[ -2\text{Ei} \Big (-\frac{1}{x}\Big)  +\text{Ei} \Big ( \frac{i\tau-1}{x}\Big) + \text{Ei} \Big ( - \frac{i\tau+1}{x}\Big)  \Big]   \Bigg\},
\end{equation}
where
\begin{gather}
    \text{Ei}(z)= -\int_{-z}^{\infty} \frac{e^{-u}}{u} du,
\end{gather}
is the exponential integral function.
\end{widetext}

\section{Quantum Coherence Measures}\label{app:Coherence}

\begin{figure}[ht]
\centering
\includegraphics[scale = 0.43]{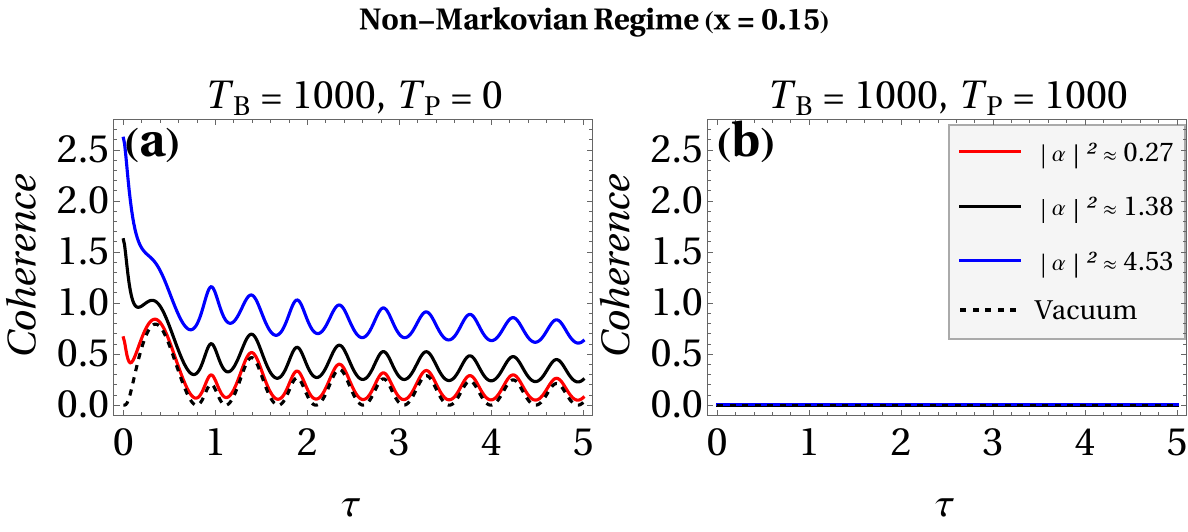}
\includegraphics[scale = 0.43]{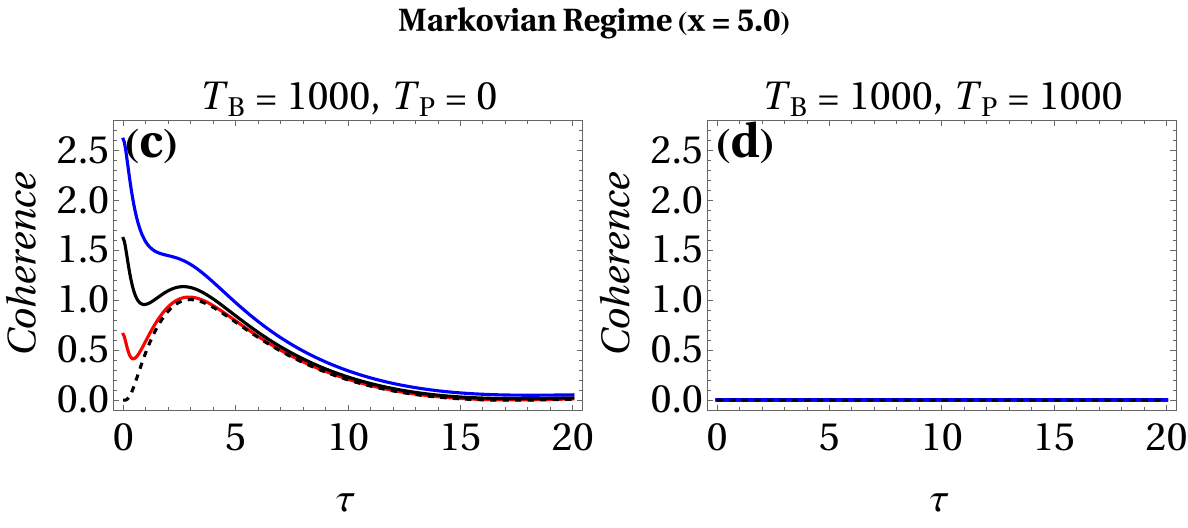}
\caption{Dynamics of relative entropy of coherence subject to diverse memory regimes and thermal conditions. Panels (a) and (b) delineate the non-Markovian regime ($x = 0.15$), while panels (c) and (d) represent the Markovian regime ($x = 5.0$). For all configurations, the bath temperature is maintained at $T_B = 1000$, and the initial squeezing is set to zero ($r=0$). Probe states are contrasted with low-temperature configurations ($T_P = 0$) presented in panels (a) and (c), and high-temperature probe conditions ($T_P = 1000$) in panels (b) and (d).  Solid curves correspond to specific symmetric displacement magnitudes calibrated to match the equivalent squeezing parameters: $|\alpha|^2 \approx 0.271$ (red), $1.381$ (black), and $4.534$ (blue) for $r=0.5$, $1.0$, and $1.5$, respectively. The reference standard classical limit ($|\alpha|^2=0$, equivalent to $r=0$) is indicated by a black dashed curve.}
\label{fig8}
\end{figure}

Within the established resource theory framework~\cite{Baumgratz2014PRL}, a quantum state $\hat{\delta}$ is classified as incoherent if its density matrix remains diagonal when represented in a specified orthonormal basis. The set of all such incoherent states is denoted as $\mathcal{I}$. Baumgratz et al. proposed measuring the coherence of an arbitrary quantum state $\hat{\rho}$ using the relative entropy of coherence:
\begin{equation}
C(\hat{\rho}) = \inf\limits_{\hat{\delta} \in \mathcal{I}} { S(\hat{\rho} || \hat{\delta}) },
\end{equation}
where $S(\hat{\rho} || \hat{\delta})= \text{Tr}(\hat{\rho}\ln \hat{\rho})-\text{Tr}(\hat{\rho}\ln \hat{\delta})$ is the quantum relative entropy. To be considered a valid coherence monotone, $C(\hat{\rho})$ must satisfy the postulates~\cite{Baumgratz2014PRL}: (\textit{i}) monotonically vanishes if and only if the state is incoherent ($\hat{\rho} \in \mathcal{I}$). (\textit{ii}) strictly non-increasing under incoherent operations, ensuring that Gaussian channels within the free operational class cannot generate coherence.
(\textit{iii}) non-increasing under the statistical mixing of quantum states.

As referenced in the main text, the proof established by Xu et al.~\cite{XuPRA2016} that single-mode Gaussian states find their optimal incoherent reference within the set of thermal states simplifies the operational quantification, allowing the analysis in terms of the probe's mean energy ($\bar{n}$) and purity or the corresponding symplectic eigenvalue ($\nu$).

In contrast to the squeezing effects explored in the main text, Figure~\ref{fig8} shows that increasing the initial probe thermal noise degrades the relative entropy of coherence for displaced states, thereby reducing their practical applicability in state-discrimination protocols over thermal-noise channels. Our methodology here is designed to address a fundamental operational question: if unitary operations (squeezing and displacement) are calibrated to have the same energy cost and initial coherence in an ideal, pure scenario, what happens to their performance when the signal source becomes noisy and thermal? Therefore, we calibrate the equivalence between the squeezing and displacement operations exclusively in the pure-state regime ($T_P = 0$). In this limit, applying a symmetric phase-space displacement of $|\alpha|^2 = \sinh^2(r)$ ensures that the purely displaced state ($r=0$) possesses the same initial quantum coherence as the corresponding pure squeezed state ($\alpha = 0$). Subsequently, to evaluate the robustness of these resources against initial preparation noise, we keep the unitary parameters ($r$ and $\alpha$) fixed and increase the initial probe temperature ($T_P \gg 0$) to observe whether the initial quantum coherence degrades due to the non-unitary addition of thermal noise. This approach directly isolates and highlights the operational resilience of the squeezing operation, demonstrating its distinct advantage over equally energetic displaced counterparts in heavily dissipative environments.

\section{Squeezing-Homodyne Receiver}\label{app:homodyne}

Here, we detail the state-discrimination strategy for a squeezing-phase-encoding protocol using continuous-variable homodyne detection. We consider the encoding of binary information into the variance profile of zero-mean Gaussian states. Assume that the sender (Alice) prepares two squeezed states with null displacement ($\alpha = 0$). The logical bit $0$ is encoded in a state $\hat{\rho}_0$ squeezed along the quadrature of position $\hat{q}$, having a narrow variance $\sigma_s^2$. In contrast, the logical bit $1$ is encoded in a state $\hat{\rho}_1$ squeezed along the quadrature of the momentum $\hat{p}$. Therefore, $\hat{\rho}_1$ must be anti-squeezed in the $\hat{q}$ quadrature, exhibiting a broader variance $\sigma_{as}^2 > \sigma_s^2$.  The receiver performs an ideal homodyne measurement of the position quadrature $\hat{q}$. The probability distributions of the continuous measurement outcomes are given by the marginal of the Wigner function integrated over the unmeasured conjugate momentum $\hat{p}$,

\begin{gather}
    P_0(q) = \frac{1}{\sqrt{2\pi\sigma_s^2}}\exp\left(-\frac{q^2}{2\sigma_s^2}\right), \\ P_1(q) = \frac{1}{\sqrt{2\pi\sigma_{as}^2}}\exp\left(-\frac{q^2}{2\sigma_{as}^2}\right).
\end{gather}
To minimize the probability of error, the optimal decision boundary must be established precisely where the two distributions are equal, $P_0(q) = P_1(q)$. Solving this equality yields two symmetric critical thresholds, $\pm q_c$:
\begin{equation}
    q_c = \sqrt{ \frac{2 \sigma_s^2 \sigma_{as}^2 \ln(\sigma_{as}/\sigma_s)}{\sigma_{as}^2 - \sigma_s^2} }
\end{equation}

\begin{figure}[hbt]
\centering
\includegraphics[scale = 0.6]{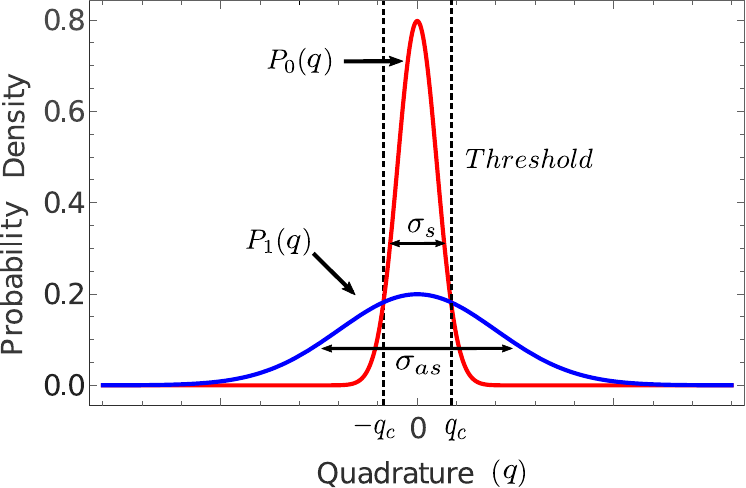}
\caption{Squeezing-homodyne receiver strategy for the discrimination of zero-displacement Gaussian states encoded in orthogonal noise quadratures. The solid curves represent the marginal probability distributions resulting from an ideal position-quadrature homodyne measurement. The position-squeezed state (bit 0) is characterized by a narrow standard deviation $\sigma_s$, while the momentum-squeezed state (bit 1) presents a broadened, anti-squeezed width $\sigma_{as}$. Optimal threshold-based discrimination is achieved by establishing decision boundaries at the critical intersection points $\pm q_c$.}
\label{fig9}
\end{figure}
If the measurement result $q$ falls within the inner region defined by the thresholds ($|q| \le q_c$), the receiver assigns the result to the position-squeezed state and declares bit $0$. If the outcome falls in the outer tail regions ($|q| > q_c$), the receiver declares bit $1$. Assuming the encoded states are transmitted with equal prior probabilities ($p_0 = p_1 = 1/2$), the error probability is computed by integrating over the distributions in their respective failure regions. The total error probability is given in terms of the error function:
\begin{equation}
    p_{\text{error}}^{\text{H}} = \frac{1}{2} \left[ 1 - \text{erf}\left(\frac{q_c}{\sqrt{2}\sigma_s}\right) + \text{erf}\left(\frac{q_c}{\sqrt{2}\sigma_{as}}\right) \right].
\end{equation}
Notably, because the encoded states are phase-space symmetric, measuring the momentum quadrature rather than the position quadrature under the same decision strategy yields an identical total error probability.

\FloatBarrier % Forces all floats before this point
%\clearpage % Ensures that all figures are processed before the references

%\bibliographystyle{apsrev4-2}
\bibliography{references}% Produces the bibliography via BibTeX.

\end{document}